# Circuit design and integration feasibility of a high-resolution broadband on-chip spectral monitor

Mehedi Hasan,[1,2,*] Gazi Mahamud Hasan,[1] Houman Ghorbani,[1] Mohammad Rad,[1] Peng Liu,[1] Eric Bernier,[2] and Trevor Hall[1]

[1]*Photonic Technology Laboratory, Centre for Research in Photonics, Advanced Research Complex, University of Ottawa, 25 Templeton Street, Ottawa, K1N 6N5, ON, Canada*
[2]*Huawei Technologies Canada, 303 Terry Fox Drive, Kanata, K2K 3J1, ON, Canada*
*\*mhasa067@uottawa.ca*

**Abstract:** Up-to-date network telemetry is the key enabler for resource optimization by a variety of means including capacity scaling, fault recovery, network reconfiguration. Reliable optical performance monitoring in general and specifically the monitoring of the spectral profile of WDM signals in fixed- and flex-grid architecture across the entire C-band remains challenging. This article describes a spectrometer circuit architecture along with an original data processing algorithm that combined can measure the spectrum quantitatively across the entire C-band aiming at ∼ 1 GHz resolution bandwidth. The circuit is composed of a scanning ring resonator followed by a parallel arrangement of AWGs with interlaced channel spectra. The comb of ring resonances provides the high resolution and the algorithm creates a virtual tuneable AWG that isolates individual resonances of the comb within the flat passband of its synthesised channels. The parallel arrangement of AWGs may be replaced by a time multiplexed multi-input port AWG. The feasibility of a ring resonator functioning over whole C-band is experimentally validated. Full tuning of the comb of resonances over a free spectral range is achieved with a high-resolution bandwidth of ∼1.30 GHz. Due to its maturity and low loss, CMOS compatible $Si_3N_4$ is chosen for integration. Additionally, the whole system demonstration is presented using industry standard simulation tool. The architecture is robust to fabrication process variations owing to its data processing approach.

## 1. Introduction

In an optical network, optical performance monitoring (OPM) is the key enabling technology for reliable spectrum management. Up-to-date network telemetry is required for capacity scaling, network or component fault recovery, and network reconfiguration through performance prediction and planning. The monitoring includes information on bit-error-rate (BER), optical signal-to-noise ratio (OSNR), electrical signal-to-noise ratio (ESNR), loss, power. The monitoring information is then passed to the network management agent for resource optimization to maximize the reach versus rate. In practice, optical performance monitoring may be based on the measurement of one or several parameters. However, OPM is used herein to refer to "power" monitoring since power is one of the key indicators of performance in optical systems.

In transport optics, especially in WDM networks, spectral sensing is not straightforward. Traditionally WDM channels are distributed over the 40 nm wide fibre-optic C-band (1530 nm to 1565 nm) with fixed centre frequencies arranged on the International Telecommunication Union (ITU) grid at intervals of 50 GHz or 100 GHz. and an OPM card is used to measure the power. The working principle of the OPM card involves sweeping a tunable filter with 50 GHz resolution over the spectrum to make available ITU grid channel power readings. Due to their excessive power consumption, size and cost, OPM cards are deployed only at a few points in the network; typically co-located with reconfigurable add-drop multiplexers (ROADMs). However, current optical networks are elastic in nature, i.e. the channels are not located on a fixed regular grid, rather the channel centre wavelength can be placed at an arbitrary location within the spectrum. The flexible grid can support a variety of channel power profiles (i.e., bandwidth and power spectral density) with a resolution of channel centre frequency placement as fine as 6.25 GHz. Flex-ready spectrum measurements are required to facilitate the deployment of the flex-grid system. As a result, flex-grid ready ROADM architectures are equipped with new modules that can measure power at desired frequency location and resolution. However, due to cost issues, spectrum measurement is only performed at add-drop nodes and not at amplifier nodes. Moreover, a single OPM module is shared by the multi degree-ROADM so the OPM measurement speed reduces as the number of lines it supports increases. Consequently, the performance of WDM channels in a section (ROADM-to-ROADM) is modeled based on an analytical or a semi-analytical analysis or a machine learning approach. The absence of OPM makes it difficult to have live and accurate network measurement; and hence it is hard to implement performance optimization. Complete knowledge of spectral content in a network is prerequisite for the effective use of color-, direction-, contention-, grid-, filter-, gap-less ROADM, flexible modulation formats, flexible channels frequencies and spectral assignment.

A variety of different approaches to the problem of spectral sensing with high resolution across a wide band have been disclosed in the literature [1-9] but when scaled to combine acceptable resolution with wideband operation their practical implementation is most often not feasible due to excessive cost, loss and footprint. An integrated solution for a high resolution (sub-GHz) spectrometer to monitor the power in fixed- and flex-grid architectures across the entire C band 1530 nm to 1565 nm remains challenging. Nevertheless, a three-stage architecture proposed in a previous publication [10] has been shown to be viable. The first stage is a tunable ring resonator (RR) that defines the resolution. The third stage is an arrayed waveguide grating (AWG) that isolates one RR resonance within each of its channels. The principal innovation is ganged tuning of the RR and AWG to retain the RR resonance at the centre of the AWG channel passband. This is achieved by a second stage that uses an MZI to form a coherent superposition of two interleaved AWG channel spectra corresponding to a pair of input ports. Further details can be found in [10]. This paper describes a refinement of the architecture and method that enables the ganged tuning of the AWG to be virtualized. The coherent superposition of a pair of AWG input channels is replaced by the incoherent superposition of pairs of AWG outputs, which may be performed by processing the measured AWG channel powers. The MZI stage is eliminated, releasing the spectrometer from any requirement to control inter-stage optical path lengths and thereby significantly easing manufacture.

In this paper, the circuit architecture design, modelling and integration feasibility of an ultra high-resolution wideband spectrometer based on the refined architecture is presented. The purpose of the proposed circuit is to measure the spectral profile of WDM channels in flex- and fixed-grid architectures across 1530 nm to 1565 nm (C band). The architecture combines a RR and a number $M$ of AWGs with interleaved channel spectra. For the purpose of exposition, the $M = 2$ case is first considered but subsequently generalised. Hardware economies may be made by replacing the multiple AWGs by a single time multiplexed $M$-input AWG. The $M = 3$ architecture has particular merit as the first in sequence to offer essentially zero crosstalk.

The proposed spectrometer can be fabricated in any mature fabrication platform. However, the resolution bandwidth of the spectrometer primarily depends on the insertion loss (dB/turn) of the RR. Hence in order to meet the specification of the proposed spectrometer, the CMOS compatible $Si_3N_4$ photonic integration platform has been selected as it offers low loss, tight confinement, low dispersion and a mature thermo-optic phase shifter technology. There are ample reports in the literature of the technological verification of spot size convertors (SSC), multimode interference couplers (MMI), tunable RR, and sub-circuits such as Mach-Zehnder interferometers (MZI) [11-13] and high port count AWG [14] fabricated using the $Si_3N_4$ integration platform. The detail simulation verification of the proposed architecture is presented using components (RR and AWG) designed for fabrication. A combination of industry standard software tools is used for the component design. A simple but novel signal processing approach enables the spectrometer to scan the entire C-band with high resolution (~1GHz) using only one dynamic control. The original signal processing method renders the proposed architecture robust to fabrication tolerances.

## 2. Proposed spectrometer circuit architecture

The schematic diagram of the proposed circuit architecture for $M = 2$ is shown in Fig. 1(a). The architecture consists of two stages; the first stage is the RR which offers the fine filtering while the last stage pair of AWGs offers the coarse filtering. In other words, the RR generates a periodic train of fine resonances spaced by its free spectral range (FSR), while the AWG isolates one RR resonance in each output channel. Hence, the output channel frequency spacing of the AWG must be equal with the FSR of the RR by design. The RR defines the spectrometer resolution bandwidth and is tunable in frequency over one FSR. Two identical AWGs with -3 dB channel passband-width close to half the output channel frequency spacing are required for the proposed circuit architecture implementation. However, the two AWGs are not driven by the same input channel (port), rather they are driven by adjacent input ports as shown in Fig. 1(a). The input channel frequency spacing is equal to half of the output channel frequency spacing (i.e. 1/2 FSR). The channel spectra of $AWG_1$ and $AWG_2$ are thereby interlaced and overlap as illustrated in Fig.1(b) for a 50 GHz channel frequency spacing. For clarity, three only of the channel spectra of each AWG are shown with possible mappings of the comb of RR resonances. In practice, 50 GHz AWGs need 88 channels to cover the whole C-band.

The comb of RR resonances is tuned by an intra-ring phase shift $\theta$. The translation in frequency of the comb is proportional to the phase shift and ranges over one FSR as $\theta$ ranges over $2\pi$ radians. For simplicity of exposition, $\theta = 0$ is taken to correspond to the alignment of the RR resonances with the $AWG_1$ channel passband centre frequencies. It follows that $\theta = \pi$ corresponds to the alignment of the RR resonances the with $AWG_2$ channel passband centre frequencies and $\theta = \pi/2, 3\pi/2$ corresponds to alignment respectively with the intersection between the upper (lower) $AWG_1$-3dB channel passband edge and the lower (upper) $AWG_2$-3dB channel passband edge.

The optical power of each output channel of both AWGs is measured by a photodetector array while the RR is scanned over an FSR under the control of the data acquisition system (DAQ). The principal novelty of the spectrometer is the construction of a virtual AWG that is tuned to retain the ring resonance within the passband of its synthesised

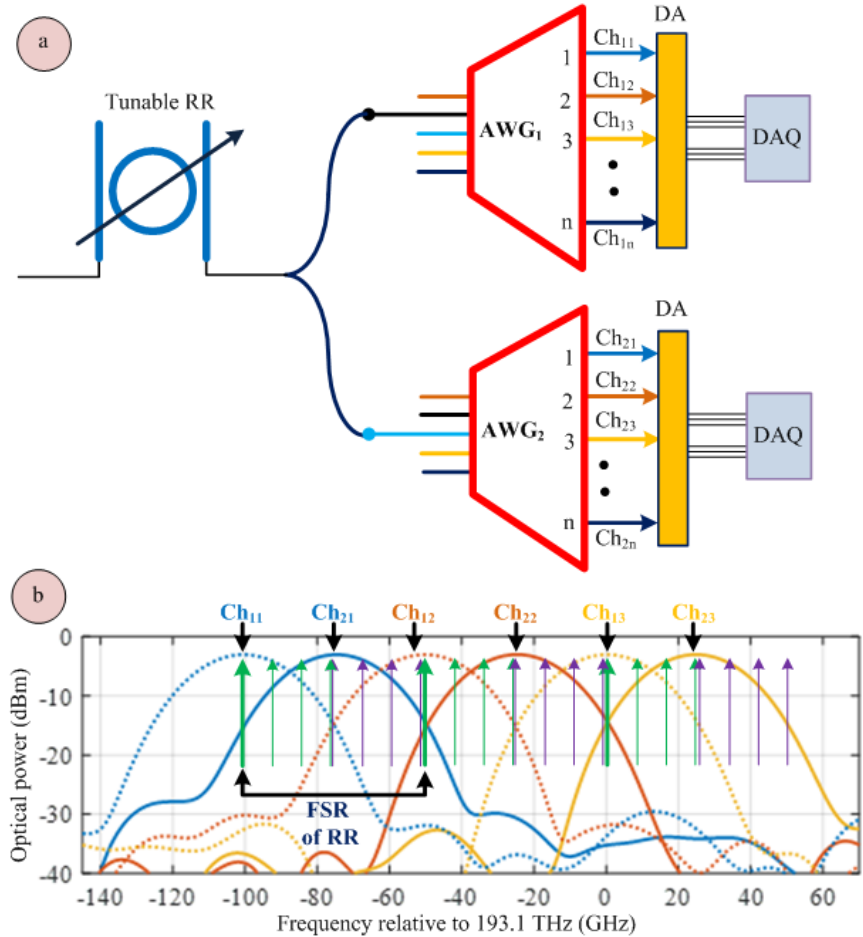

Fig. 1. (a) Schematic of the proposed spectrometer; (b) interlaced optical spectrum of $AWG_1$ and $AWG_2$ with the resonance mapping over one FSR. AWG, arrayed waveguide grating; RR, ring resonator; DA, detector array; DAQ, data acquisition Ch; channel; FSR, free spectral range.

channels A virtual channel with index $m$ is synthesised by summing the optical power of a selected pair of interlaced $AWG_1$ and $AWG_2$ channels in two phases:

**Phase 1:** As the ring resonance is tuned over the first half of the FSR by a tuning phase shift from $\theta = 0$ to $\theta = \pi$ (green stem in Fig. 1(b)), $AWG_1$ channel $m$ ($Ch_{1m}$) and $AWG_2$ channel $m$ ($Ch_{2m}$) are summed.

**Phase 2:** As the ring resonance is tuned over the second half of the FSR by a tuning phase shift from $\theta = \pi$ to $\theta = 2\pi$ (purple stem in Fig. 1(b)), $AWG_2$ channel $m$ ($Ch_{2m}$) and $AWG_1$ channel $m+1$ ($Ch_{1(m+1)}$) are summed.

If the AWG channels have a raised cosine spectral profile, then the synthesised channels have a perfectly flat passband. In practice the AWG channel profile will differ from the ideal and there will be ripple. However, the ripple will be small as the synthesised channel spectral profile will agree with the ideal when the ring resonance is aligned at either passband centre or at the intersection of the -3dB passband edges of the AWG channels summed.

**Table 1. Data processing algorithm for a 2 AWG architecture**

| RR tuning phase | Synthesised channel | Remarks |
|---|---|---|
| $\theta \in [0, \pi]$ (1st half FSR) | $Ch_{11} + Ch_{21}$ <br> ⋮ <br> $Ch_{1n} + Ch_{2n}$ | green stem Fig. 1(b) |
| $\theta \in [\pi, 2\pi]$ (2nd half FSR) | $Ch_{21} + Ch_{12}$ <br> ⋮ <br> $Ch_{2(n-1)} + Ch_{1n}$ | purple stem Fig. 1(b) |

Table 1 tabulates the steps of the data processing algorithm performed in the electronic domain by the data acquisition system. Handover from phase 1 of the processing algorithm to phase 2 occurs at $\theta = 180°$ but this is not critical. Handover may occur for $\theta$ anywhere between 165° or 195° with very little or no penalty. Hence, the careful tracking of the position of a resonance is not mandatory. However, an integrated wave meter concept presented in [16] can be used to monitor the position of the RR resonance within a single channel of the AWG. Since the resonances are substantially periodic in frequency, albeit very slightly detuned by chromatic dispersion, the position of the resonance within each channel of the AWG can be mapped easily.

The spectrometer requires only one control, which sets the intra-ring phase shift in order to tune the RR resonant frequency comb cyclically over one FSR. Our device and circuit simulations; previously reported experimental demonstrations; and the process development kit support the practicality of a 50 GHz FSR RR. Table 2 shows the detail specifications of the proposed circuit design. For a RR FSR of 50 GHz, the required number of AWG output ports is 88 to cover the entire C-band.

**Table 2. Detail design specifications of the proposed circuit architecture shown in Fig. 1(a)**

| Design specifications | Numbers | Remarks |
|---|---|---|
| RR free spectral range (GHz) | 50 | |
| AWG output channel spacing (GHz) | 50 | equal to the free spectral range of RR |
| AWG output channel bandwidth (GHz) | 20 | ∼ ≤ 1/2 AWG output channel spacing |
| AWG input channel spacing (GHz) | 25 | half of AWG output channel spacing |
| Number of AWG output channels | 88 | |
| Total spectrum covered (THz) | 4.4 | 88 × 50 = 4400 GHz (entire C band) |

The number of output ports can be reduced by increasing the FSR of the RR, since 50→150 GHz channel spacing AWGs having output ports up to 96 are available commercially [15]. On the other hand, a RR having FSR of 220 GHz fabricated using double strip TriPleX™ waveguide technology is already reported in [3]. Hence, the proposed number of output channels of the AWGs can be reduced to 1/3 (∼30 channels) of the 50 GHz example.

The VPIphotonics simulation tool is used to evaluate the performance of the combined circuit architecture and data processing method. The optical power at each output channel of the both AWGs is monitored while scanning the RR over one FSR. The calculation of the power depends on the number of samples used in the simulation time window. Since the signals were widely spread, a moderate time window is used in the simulation to avoid an excessive memory requirement. A slight variation in the result is obtained for different time window settings. The simulated output is then processed using the algorithm presented in Table 1. Figure 2 shows the variation in measured optical power as a function of the resonance position over one FSR for various AWG channel passband widths. Channel 1 & 2 of $AWG_1$ and channel 1 of $AWG_2$ are used for the calculation. The results show that the spectral measurement is almost flat with little ripple (∼0.5 dB) over the entire FSR for a passband width of 25 GHz. The ripple increases inversely with the AWG passband and reaches ∼1.25 dB for a passband width of 20 GHz. Figure 2(c[i-iv]) shows the optical spectrum of $AWG_1$ and $AWG_2$ at output channel one ($Ch_{11}$ & $Ch_{21}$) for ring resonance position of 0° and 90° respectively for a passband width of 25 GHz. It shows that a strong (∼-21→-22 dB) crosstalk component is present at the $AWG_2$ channel. Hence, the power calculation is misleading at the 0° position. The crosstalk component can be reduced to ≤-30 dB by reducing the passband width of the AWG. Figure 2(c[v-viii]) shows the optical spectrum of $AWG_1$ and $AWG_2$ at the same resonance position for the pass band width of 20 GHz. The crosstalk at the neighboring channel is reduced by ∼-8→10 dB. This leads to a ripple of ∼1→1.25 dB over the entire FSR. Since the transmission characteristic of an AWG is substantially periodic, the simulated result will almost be same for any other channel measurement of the AWG. The entire channel spectra of both AWGs are presented in Fig. 8.

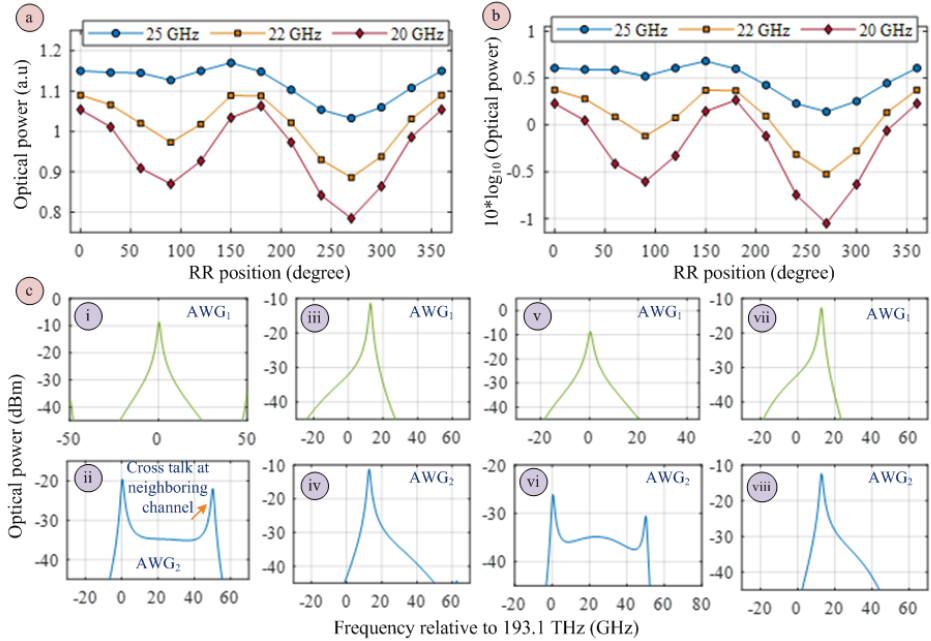

Fig. 2. Simulated optical power as a function of ring resonance for different passband width of the AWG channel, (a) linear scale, (b) logarithmic; (c) optical spectrum of $AWG_1$ and $AWG_2$ at channel one for ring resonance position of 0° and 90° for the passband width of 25 GHz [i-iv] and 20 GHz [v-viii] respectively.

To substantially eliminate adjacent channel crosstalk and to reduce the ripple, the architecture is upgraded to three AWG with interlaced channels as shown in Fig. 3(a). For simplicity, assume the three AWGs are identical with input port spacing of $1/3$ of the output channel spacing, with $AWG_1$, $AWG_2$ and $AWG_3$ driven by input port $l$, $(l+1)$ and $(l+2)$ respectively; where $l$ is an integer. Figure 3(b) shows the interlaced spectrum of the upgraded architecture using the parameters given in Table 3.$(l+2)$ respectively; where $l$ is an integer. Figure 3(b) shows the interlaced spectrum of the upgraded architecture using the parameters given in Table 3.

Table 3. Detail design specifications of the proposed circuit architecture shown in Fig. 3(a).

| Design specifications | Comments | Remarks |
| --- | --- | --- |
| RR free spectral range (GHz) | 51 | |
| AWG output channel spacing (GHz) | 51 | equal to the free spectral range of RR |
| AWG output channel bandwidth (GHz) | 17 | ∼ ≤ 1/3 AWG output channel spacing |
| AWG input channel spacing (GHz) | 17 | 1/3 of AWG output channel spacing |
| Number of AWG output channel | 88 | |
| Total spectrum covered (THz) | 4.4 | 88 × 51 = 4488 GHz (entire C band) |

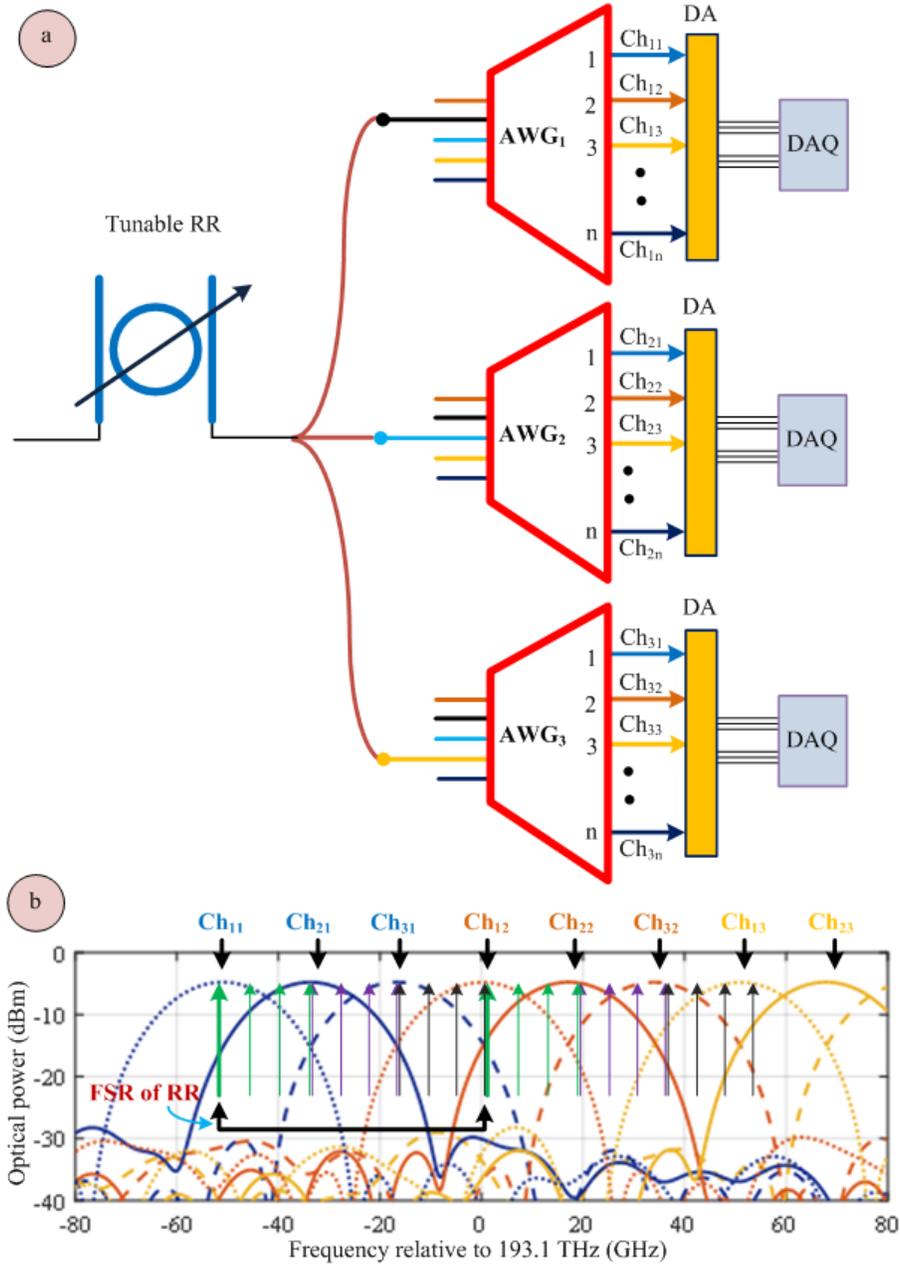

Fig. 3. (a) Schematic of the modified design; (b) interlaced spectrum of AWG$_1$, AWG$_2$ and AWG$_3$ with the resonance mapping over one FSR. The AWG channel profile is defined using VPIphotonics's built-in physical model.

A virtual channel with index $m$ is synthesised by summing the optical power of a selected pair of interlaced AWG$_1$, AWG$_2$, AWG$_3$ channels as before but now in three phases:

**Phase 1:** As the ring resonance is tuned over the first third of the FSR by a tuning phase shift from $\theta = 0$ to $\theta = 2\pi/3$ (green stem in Fig. 3(b)), AWG$_1$ channel $m$ (Ch$_{1m}$) and AWG$_2$ channel $m$ (Ch$_{2m}$) are summed.

**Phase 2:** As the ring resonance is tuned over the second third of the FSR by a tuning phase shift from $2\pi/3$ to $\theta = 4\pi/3$ (purple stem in Fig. 3(b)), AWG$_2$ channel $m$ (Ch$_{2m}$) and AWG$_3$ channel $m$ (Ch$_{3m}$) are summed.

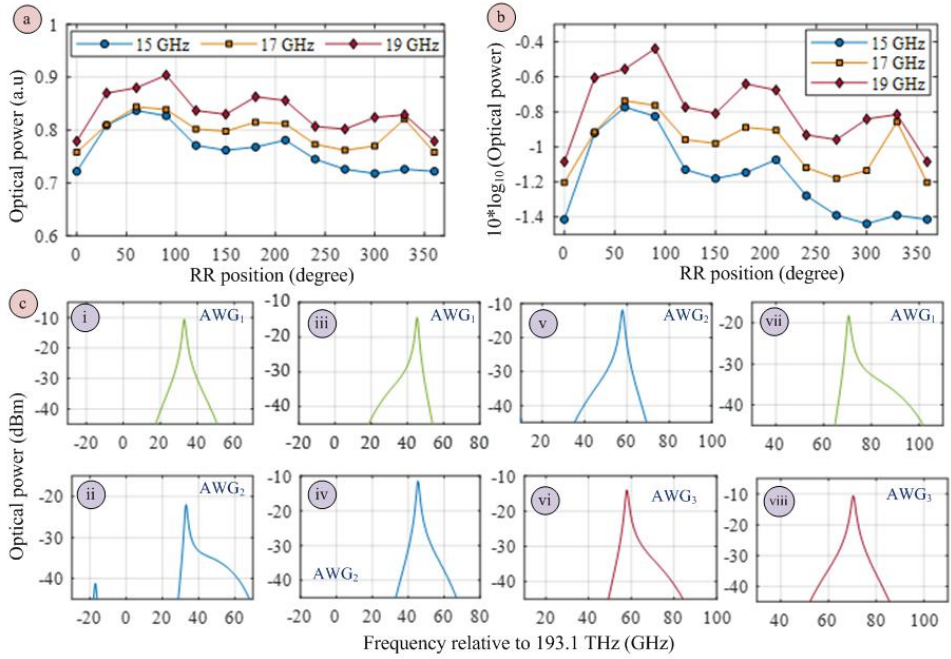

Fig 4. Simulated optical power as a function of ring resonance for different passband width of the AWG channel (a) linear scale (b) logarithmic scale; (c) optical spectrum of $AWG_1$, $AWG_2$ and $AWG_3$ for ring resonance position of 0˚[i-ii], 90˚[ii-iv], 180˚ [v-vi] and 270˚[vii-viii] for the passband width of 17 GHz.

**Phase 3:** As the ring resonance is tuned over the final third of the FSR by a tuning phase shift from $\theta = 4\pi/3$ to $\theta = 2\pi$ (black stem in Fig. 3(b)), $AWG_3$ channel $m$ ($Ch_{3m}$) and $AWG_1$ channel $m + 1$ ($Ch_{1(m+1)}$) are summed.

Table 4 tabulates the steps of the data processing algorithm performed in the electronic domain by the data acquisition system. Figure 4 shows the variation in measured optical power as a function of the resonance position over one FSR for three AWG channel passband widths. The ripple in the spectrum sensing is reduced to $\leq 0.4$ dB for the preferred passband width $1/3$ (17 GHz) of the output channel spacing (51 GHz). Due to fabrication tolerances, the passband width may vary from 17 GHz, as such, two other measurements are taken for width of 15 GHz and 19 GHz respectively for all three AWGs. Almost identical results are obtained. Ideally, the measured optical power should be constant as there is negligible crosstalk present in the design. The variation obtained in the simulation is an artifact to the time window settings. Fig. 4(c[i-viii]) shows the optical spectrum at different location of the ring resonance.

**Table 4. Data processing algorithm for a 3 AWG architecture**

| RR tuning phase | Synthesised channels | Remarks |
|---|---|---|
| $\theta \in [0, 2\pi/3]$ (1st third FSR) | $Ch_{11} + Ch_{21}$ <br> ⋮ <br> $Ch_{1n} + Ch_{2n}$ | green stem Fig. 3(b) |
| $\theta \in [2\pi/3, 4\pi/3]$ (2nd third FSR) | $Ch_{21} + Ch_{31}$ <br> ⋮ <br> $Ch_{2n} + Ch_{3n}$ | purple stem Fig. 3(b) |
| $\theta \in [4\pi/3, 2\pi]$ (3nd third FSR) | $Ch_{31} + Ch_{12}$ <br> ⋮ <br> $Ch_{3(n-1)} + Ch_{1n}$ | black stem Fig. 3(b) |

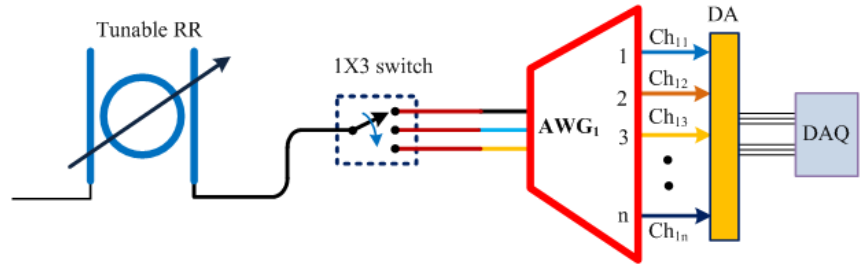

Fig. 5. Schematic diagram showiung practical implementation.

There is no crosstalk present in the spectrum and the spectral sensing is very simple and straightforward. Fig. 4(c[i-iv]) shows the spectrum of $AWG_1$ and $AWG_2$ for resonance position $\leq 120°$. Figure 4(c[v-vi]) shows the spectrum of $AWG_2$ and $AWG_3$ obtained at resonance position $180°$. Finally, Fig. 4(c[vii-viii]) depicts the optical spectrum of $AWG_3$ and $AWG_1$ measured at the resonance position of $270°$. Since the overall spectrum is delivered to all the three AWG but the measurement involves summing the outputs of only two AWG, $1/3$ of the total input power is discarded. Furthermore, it requires one more AWG over the scheme presented in Fig. 1(a). The huge advantage of the design is spectral sensing is performed with essentially no adjacent channel crosstalk.

The practical implementation can be made simpler by time multiplexing using a single 3-input AWG and a $1 \times 3$ switch as shown in Fig. 5. The input ports are spaced in frequency by $1/3$ of the AWG output frequency spacing. For simplicity, assume the resonance position is at $0°$. The switch is connected to the upper input port of the AWG. The output data is recorded by the DAQ. Now change the switch to the middle input and record the data. Repeat the procedure for resonance position up to $120°$. Alternatively; keeping the switch at upper input port, scan the ring resonance up to $120°$ and record the data. Now change the switch to middle input port, scan the ring resonance from $0° \rightarrow 120°$ and record data. Now data processing will be done as per Table 4. For resonance position $120° \rightarrow 240°$, the switch will be toggled between the middle input and last input port of the AWG. Finally, for $240° \rightarrow 360°$, the switch will toggle between 1st input and third input port of the AWG. As before there is very little penalty in the measured optical power if handover between phases occurs within $\pm 20°$ of the specified tuning phases of $120°, 240°, 360°$. The switch does not need to toggle at the exact positions specified and the data acquisition can be performed flexibly.

## 3. Validation & integration feasibility

The feasibility of the ring resonator is verified experimentally whereas the whole system is demonstrated using industry standard simulation tool. A combination of software simulation tools is used to validate the concept. For example, VPIphotonics is used for the circuit simulation whereas, the Photon Design suite of tools and OptiBPM are used to verify the function of each component that comprise the architecture. The proposed spectrometer can be fabricated in any mature low-loss photonic integration platform. If the excess loss of the ring per turn is negligible in comparison to the power coupled out per turn, the resolution bandwidth of the spectrometer is determined by the power cross-coupling ratio of the couplers. The ring excess loss per turn consequently limits the achievable resolution. An integration platform supporting the design of low-loss waveguides and low-loss waveguide bends is therefore paramount. Owing to its low loss, tight confinement, low dispersion waveguides and a mature thermo-optic phase shifter technology, the CMOS compatible $Si_3N_4$ photonic integration platform is selected to meet the specification of the proposed spectrometer circuit. The platform also offers good prospects for further loss-reduction [17] and to lower power consumption, temperature insensitive, alternatives to thermo-optic phase-shift elements [18-19]. For effective use of resources, the original plan for fabrication envisaged access to multi-project wafer MPW runs for test structures followed by a custom wafer run for fabrication of prototypes for demonstration. LioniX only offer $Si_3N_4$ Triplex technology MPW runs for designs using the asymmetric double strip (ADS) waveguide and the low-cost photolithography used has a minimum feature size of 1 $\mu$m. Accordingly, the simulations of the components and sub-circuits that constitute the proposed spectrometer are designed using ADS as reference waveguide. The waveguide characteristic over full C-band is obtained by using the Photon design software tool FIMMWAVE. TE-like mode is used in all the simulation due to its tight confinement, hence it exhibits lower bend loss in comparison to the TM-like mode. The effective group index of the mode at the at the smaller wavelength edge (1530 nm), centre wavelength

(1545 nm), and the longer wavelength edge (1565 nm) of the C-band is found to be 1.7725, 1.76841 and 1.7629 respectively. Figure 6(a) shows the schematic diagram of a RR. As shown, it requires two directional couplers (DC). The resolution of the spectrometer is largely set by DC power cross coupling and is determined by the spatial gap between the interacting waveguides. A variety of numerical and quasi analytical methods involving different approximations are used to bracket a range of gaps targeting 2-6% power coupling. The rings with FSRs of the order of 50 GHz fall into a simulation no man's land. 3D FDTD requires too large computational resources. 2.5D FDTD uses effective index mode (EIM) solvers which cannot correctly model couplers. The eigenmode expansion (EME) method has problems modelling curved structures – it can be used in conjunction with a circuit simulator (PICWAVE, VPI) for very small FSR large racetrack rings but it displayed an excessive computational power loss in the curved waveguide coupler of the 50 GHz ring. In this work, a semi-analytic method informed by a mode solver was applied to the problem. The mode solver tool (Photon Design MOLAB) is applied to find the fundamental symmetric and antisymmetric local eigenmodes to provide their effective index difference which is integrated over the interaction region using an adiabatic symmetry-based model that predicts the overall power transfer matrix of the proximate curved waveguides. An OptiBPM scanning of the power coupling as shown in Fig.6 (b) is also performed to validate the numerical calculation. Although the beam propagation method falls outside its domain of validity for this problem, it nevertheless predicted gaps in a similar range as the quasi-analytic method. The range of gaps is then sampled by test structures to enable the gap to be refined experimentally. The zoom in view of the power coupling at the cross port of the DC is shown in Fig. 6(c). Simulation result show that, for a gap of 1.3 um the power coupling at the cross port is 5% and it reduces to <~1% for a gap of 1.7 um. For the MPW run, several ring resonators are laid out on the mask having gaps from 1.2 um to 1.8 um with 0.2 um increment with the objective that at least one RR works well. Two identical DC are used in the ring design. The ADS waveguide ring circumference is calculated to be 3.3928 $mm$ at 1545 nm for an FSR of 50 GHz using the following formula.

$$FSR(\omega) = c/n_g(\omega)l \tag{1}$$

where, $n_g$ is the group index, $l$ is the length of the delay line to obtain the specified FSR. The corresponding ring radius is 539.99 $\mu m$. The bend loss of an ADS waveguide bend of this radius of curvature is negligible over the C-band; the mode is fully bound by the waveguide bend and the only terms contributing loss are absorption and scattering loss. Detailed data on absorption and scattering loss is not available beyond the disclosure that the total waveguide loss is circa 0.1-0.2 dB cm$^{-1}$ for a ring of 50-100 $\mu m$ radii. The waveguide loss is considered as 0.4 dB/cm in the simulation as a conservative assumption. A detailed ring response can be found in [10] using VPIphotonics simulation. Figure 7 (a-c) shows the measured transmission spectra of the RR at various point of the C-band. The peak transmission of the RR varies by 1.25→ 1.75 dB from the longer wavelength edge to the shorter wavelength edge of the C- band. Fig. 7(d) shows the zoom-in view of the ring resonance at the design wavelength. The full width half maximum (FWHM) bandwidth of the RR is found to be around ~1.3 GHz. Figure 7 (e-f) shows the measured FSR at the centre wavelength and longer wavelength edge. Furthermore, the FSR is ~48.70 GHz at the edge of the short wavelength. The FSR variation is due to the dispersion of the group index across the band from the index at the design wavelength, 1545 nm. Further, digital signal processing can be used to provide the quantitative spectrum. The LioniX MPW fabrication process offers only thermo- optic phase shifts. Hence, the tuning of the RR is done using such a phase shifter. Figure 8 (a) shows the tuning of the ring resonances as a function of applied voltage to the thermo-optic phase shifter. A full FSR tuning is obtained. Figure 8 (b) shows the I-V characteristic of the heater. The direct measurement of the phase shifter resistance is found to be 734 Ω; well aligned with the slope (~730 Ω) of the I-V curve. While the peak resonances as a function of heater is shown in Fig. 8(c). As shown the tuning wavelength (frequency) is proportional to the applied heater power. However, due to low drive voltage, low drive power and linear voltage to index relationship, electro-optic tuning offers smooth operation with better efficiencies [20]. As an alternative means of tuning, the insertion of piezo-electric actuator augurs well for the future [19].

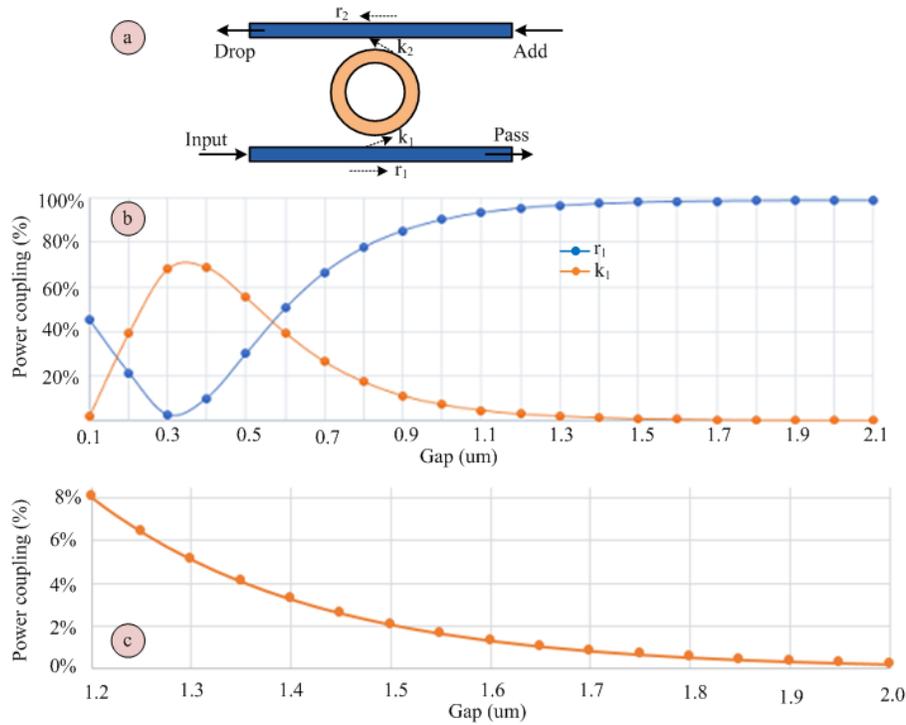

Fig. 6. (a) Schematic diagram of a RR; (b) power coupling between the ports of a directional coupler as a function of spatial separation among them; (c) zoom-in view of the power coupling at the cross port as a function of spatial separation.

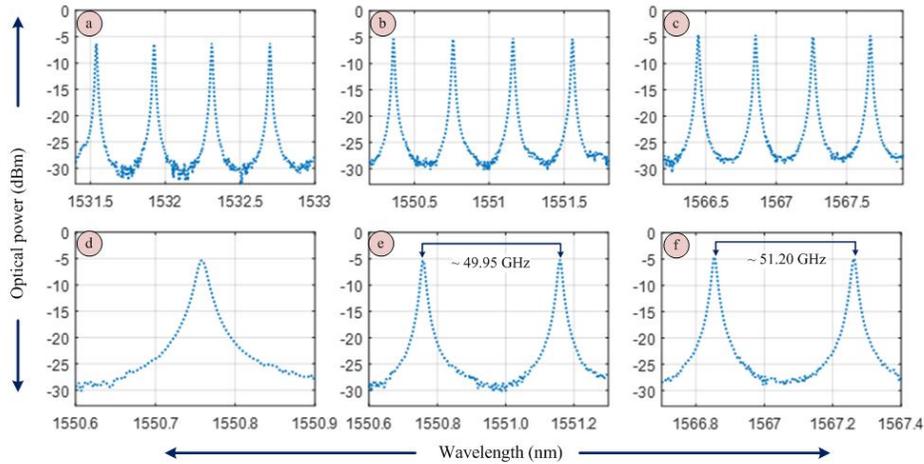

Fig. 7. Measured transmission spectrum of the RR; (a) smaller wavelength edge; (b) center wavelength; (c) longer wavelength edge; (d) zoom-in view of the ring resonance at the design wavelength. The measured full width half maximum (FWHM) is ~ 1.30 GHz; FSR at the (e) center wavelength and; (f) longer wavelength edge.

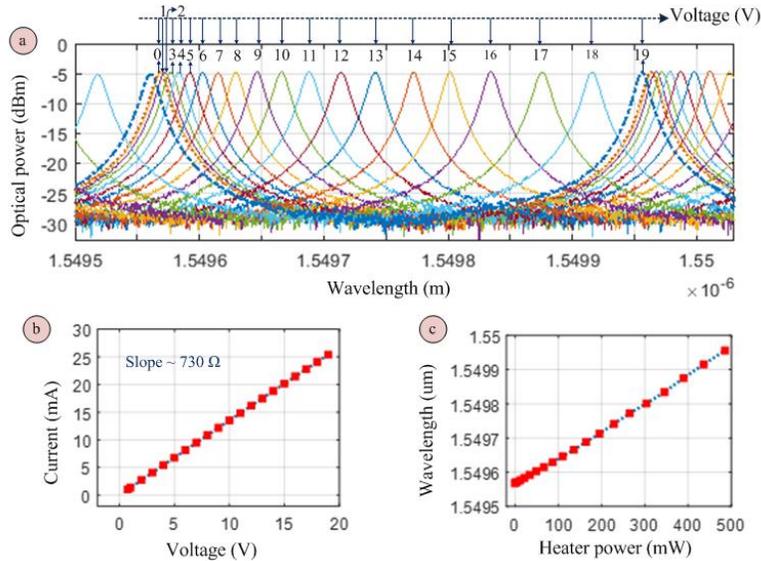

Figure 8. (a) Tuning of the ring resonance as a function of applied voltage to the thermo-optic phase shifter; (b) The I-V characteristic of the thermo optic phase shifter and (c) Peak resonances in a particular FSR as a function of heater power.

Table 5. Design specifications detail of the $AWG_1$ and $AWG_2$.

| Design specifications | Comments | Remarks |
|---|---|---|
| AWG output channel spacing (GHz) | 50 | equal to the free spectral range of RR |
| AWG output channel bandwidth (GHz) | 20 | ~ ≤ 1/2  AWG output channel spacing |
| AWG input channel spacing (GHz) | 25 | half of AWG output channel spacing |
| Number of input channel | 8 | |
| Number of output channel | 32 | |
| Free spectral range FSR (GHz) | 1600 | 32×50 GHz |

While the fine tuning of the RR is performed, the AWGs perform the coarse filtering, that is isolating each resonance of the RR at the output ports of the AWGs. The AWG design is performed by Bright Photonics BV to the specification given in Table 5. Two identical AWGs except for a channel spectrum shifted by 25 GHz have been designed. Figure 9(a) shows the designed transmission spectra of the upper AWG as a function of frequency. As shown in the table 5, a 32-channel cyclic AWG is designed for fabrication. Two AWGs ($AWG_1$ & $AWG_2$) are identical in design apart from a 25 GHz relative shift of the channel spectra. Figure 9(b) shows a zoomed-in view of the interlaced $AWG_1$ and $AWG_2$ channel spectra when light is launched from the input channel one. It also shows that the crossover between the channels are down to ∼-22 dB, this was the main aim of the reduction of the passband width to 20 GHz. Similar results were obtained when light is launched from other input ports. For a ring resonator having FSR of 50 GHz, an AWG of 86∼88 channels are required to cover the whole spectrum. The 32- channel AWG design presented here can easily be scaled up to 86 or 88 channels since a higher number (96) of output channels commercial AWG is already available [15]. Since the designed 32-channel AWG is cyclic, the output spectrum is periodic with a period of 32. Hence, with the aid of a tunable optical bandpass filter the spectrum of the whole C-band can be measured to prove the concept. Figure 10 shows the zoom-out spectra of the designed AWG. It shows that the envelope of the AWG passband is ∼1.8 dB down at the edges of the AWG FSR compared to the centre of the AWG FSR. This can be reduced significantly by increasing the FSR of the AWG.

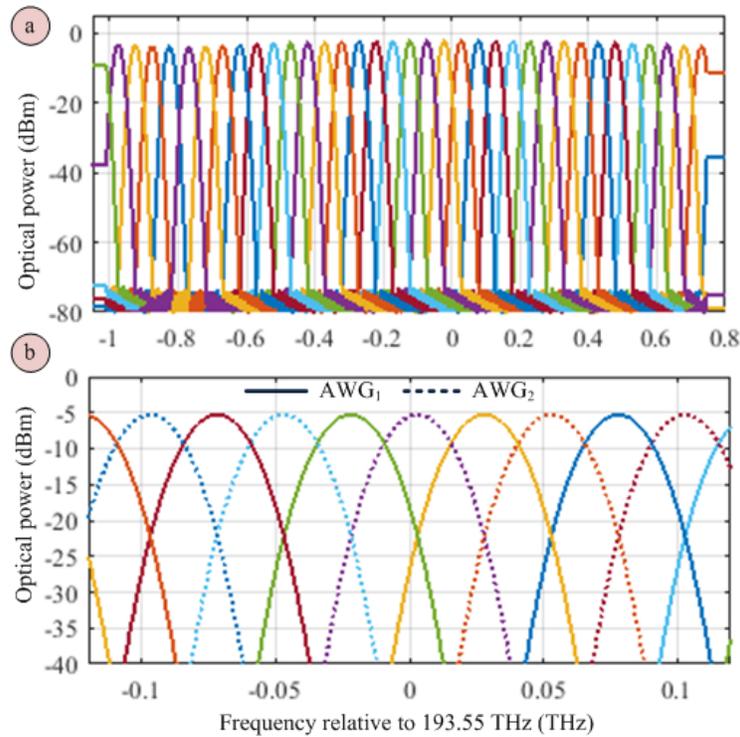

Fig. 9. (a) Simulated spectrum of the $AWG_1$ with input light launch from upper left port (port 1); (b) zoom-in view of the interlaced spectrum between $AWG_1$ and $AWG_2$. The input light is launched from the identical input port (port 1).

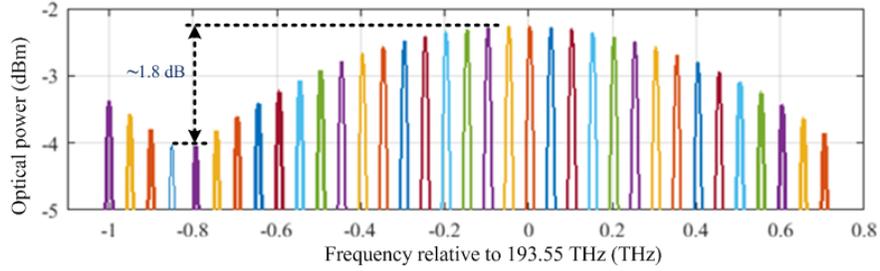

Fig. 10. Transmission spectra of the AWG output-channels. Zoom-out view of Fig. 8(a).

The circuit level simulation has been implemented using VPIphotonics to evaluate the overall performance. The RR is configured similar to Fig. (7). as explained in [10]. The S matrix of the AWGs provided by Bright Photonics BV is imported into VPIphotonics accordingly. The scanning is performed by changing the position of the ring resonance across one FSR while recording the available optical power at the PIN diode at the output of each AWG channel. The arithmetic summation is performed offline as per Table 1 to obtain the plot as shown in Fig.11 (a-b) and 12 (a-b). The results presented in Fig. 11 are obtained when both the AWGs are driven from input channel one (1). Whereas, Fig. 12 is obtained by launching the light from input channel four (4). Both the figures are almost identical. Fig. 11 (a-b) shows the available optical power at the PIN photodiode over one complete FSR. Figure 11 (a) represents the optical power variation as a function of frequency on a linear scale, whereas Fig. 11 (b) uses a semi log scale. The optical power variation is plotted on a logarithmic scale. The measured power at the edges of the FSR is ∼1.5→2 dB less than the power at the centre of the FSR. This finding is consistent with the envelop of the AWG passband reported in Fig. 10. The inset shows a zoomed-in view of the plot at three different resonance locations within the FSR. It shows that a ripple of ∼1→1.25 dB is present in the detected power; this is identical with the findings presented in Fig. 2(b) obtained by simulation for the ideal case. Figure 11 (c) shows the optical spectrum obtained from two different channel located at the edges (left and right) of the AWG FSR for two different position of the RR. The spectrum presented in the upper row is obtained at the output channel of $AWG_1$ while the spectrum in the lower row is obtained from $AWG_2$. Figure 11 (c) (i-iv) is obtained for the RR position of 0°. The position 0° refers to the alignment of the ring resonances with the passband centre of the AWG. This is taken as the reference (starting) point to simplify the calculation. Whereas spectrum presented in Figure 11 (c) (v-viii) obtained for ring position of 90°; slightly to the right of the origin at 0°. Figure 11 (c) (ii, iv) shows that the crosstalk contribution in the neighbouring channel is reduced greatly by reducing the AWG passband bandwidth to 20 GHz. The simulated crosstalk at the neighbouring channel is less than < −30 dB for the worst cases. Figure 12 is identical to Fig. 11 albeit the light is launched from input channel four (4). In addition, the spectrum presented in Figure 12 (c) (v- viii) is for the RR position of 120° rather than for 90° in Fig. 11.

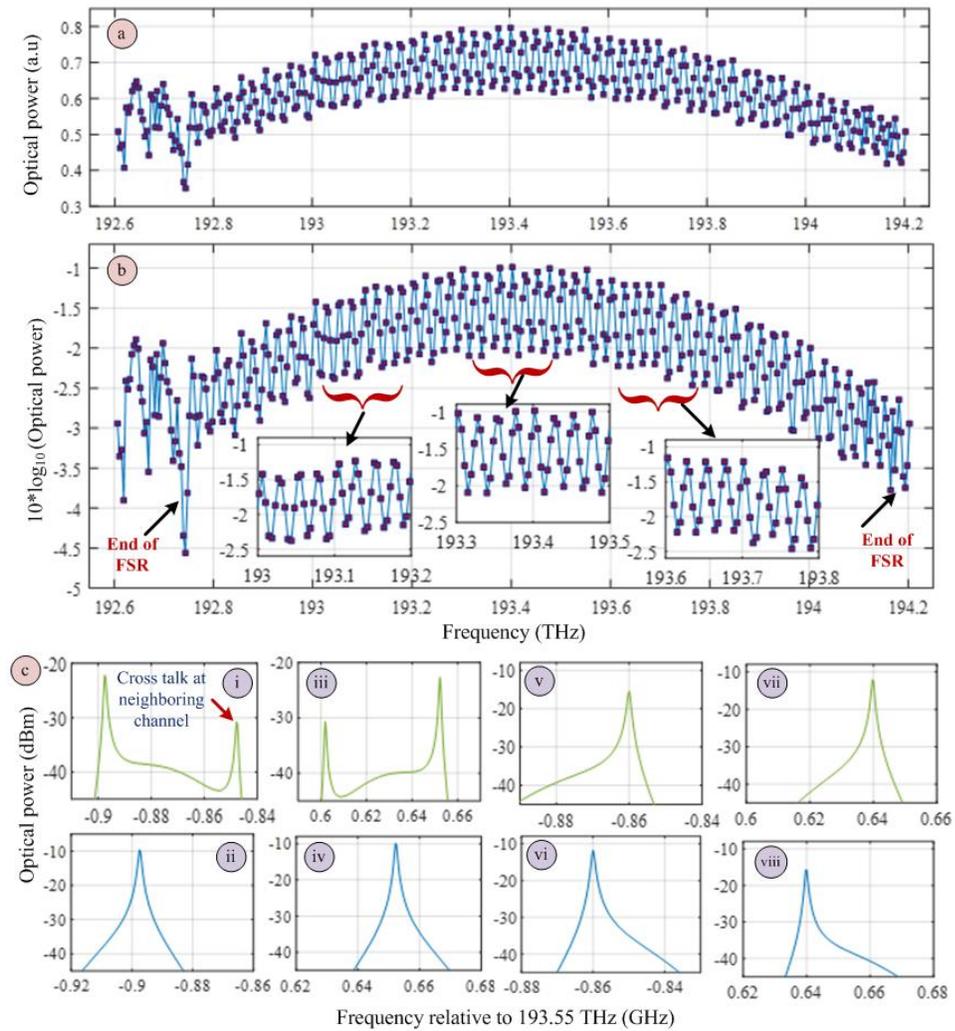

Fig. 11. Simulated result of the whole circuit when light is launched from input channel one (1) of both AWG; (a) optical power available at the PIN photodiode as a function of frequency; (b) logarithmic representation of the optical power measured in (a); (c) optical spectrum at the left (i-ii & v-vi) and right edges channel (iii-iv & vii-viii) of the AWG FSR for 0° and 90° degree position of the RR respectively. Upper row represents the spectrum of $AWG_1$ while lower row represents the spectrum of $AWG_2$. The spectrum sensing is obtained in (a) by simply summing the detectable optical power in $AWG_1$ and $AWG_2$. In this case (i)+(ii) for the frequency offset ∼-0.89 THz and (iii)+(iv) for frequency offset ∼0.65 THz from the center (193.55 THz) of the AWG FSR.

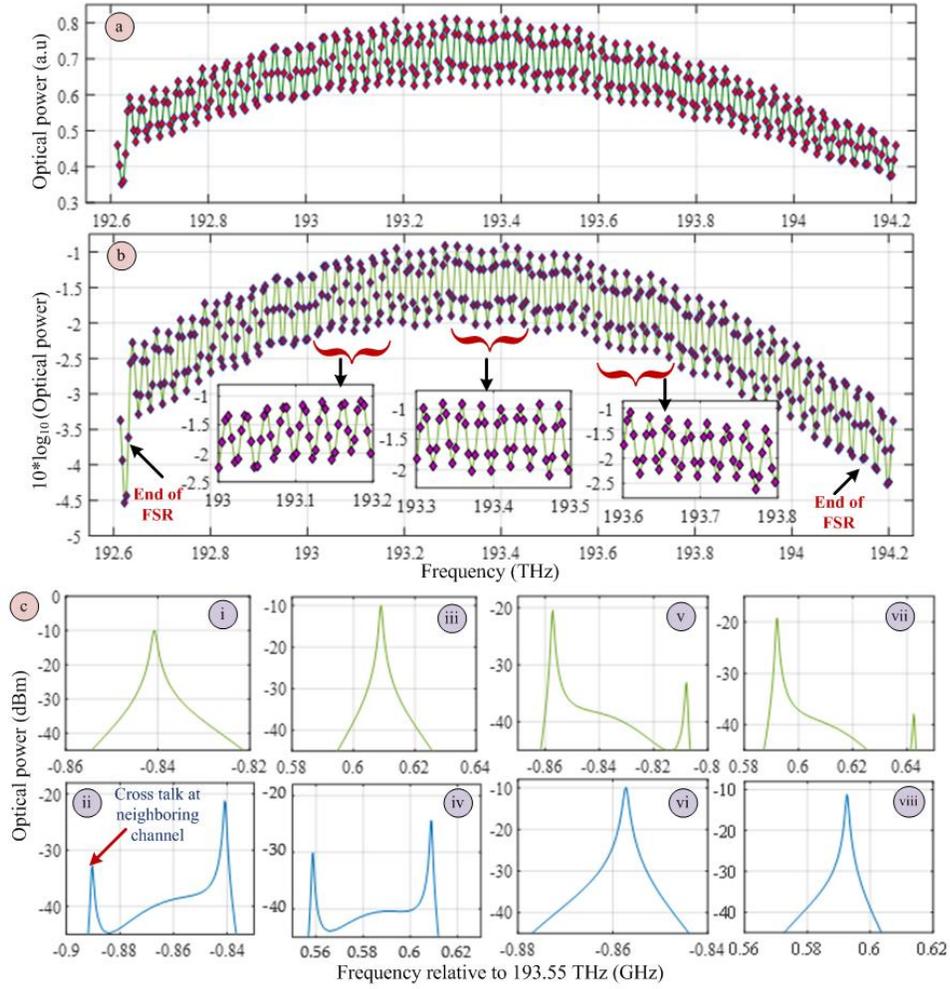

Fig. 12. Simulated result of the whole circuit when light is launched from input channel four (4) of both AWG; (a) optical power available at the PIN photodiode as a function of frequency; (b) logarithmic representation of the optical power measured in (a); (c) optical spectrum at the left (i-ii & v-vi) and right edges channel (iii-iv & vii-viii) of the AWG FSR for 0˚ and 120˚ degree position of the RR respectively. Upper row represents the spectrum of $AWG_1$ while lower row represents the spectrum of $AWG_2$.

## 4. Discussion

For simplicity of understanding, no detuning of the ring resonance from the AWG channel centre frequencies at the limits of the scan is assumed in the earlier sections. This assumption is not necessary since the scanning is performed based on the simple arithmetic summation of the output of AWGs based on Table 1. A wave meter [16] can be placed at an outer channel to monitor the position of the ring resonance within the channels. The channel shift between the $AWG_1$ and $AWG_2$ should be half (25 GHz in this case) of the AWG output channel spacing; is the only restrictions of the design. If required, one of the AWG can be temperature tuned to obtaining the required frequency shift.

Figure 13 shows the impact of the frequency shift between the AWG. When the AWGs are shifted precisely (25 GHz), the measured optical power is symmetrical about the RR position. However, when the channels are slightly detuned from the exact shift, the symmetry is broken. One half of the RR tuning produces more ripple, whereas the ripple is reduced significantly in the other half of the RR tuning. One can adopt simple signal processing techniques to eliminate this asymmetry. For a 21 GHz frequency shift between $AWG_1$ and $AWG_2$ channels, the output power manifests ~2 dB additional ripple. In the AWG design the adjacent crosstalk is found to be more than 25 dB. In practice, it may be well below 25 dB. In that case, the intensity reading of a particular channel is misleading as the detector produces the sum of all signal components. If the AWG channel characteristic (adjacent crosstalk) is known, the actual intensity can be obtained by subtracting the intensity contribution due to the adjacent channel crosstalk. This simple processing can be done after data acquisition. The adjacent channel crosstalk mainly depends on the phase errors in the arrayed waveguide sections. A ring resonator having FSR of 220 GHz using $Si_3N_4$ has already been reported in [3]. For an FSR of 200 GHz, the AWG can be reduced to 22 channels in order to cover the whole C-band. This makes the AWG design compact and more robust to process variations. For example, an adjacent crosstalk of ∼-22 dB is obtained for a 16 × 16 cyclic AWG using high contrast silicon photonics having 189 GHz output channel spacing [21]. On the other hand, an adjacent crosstalk of -18 dB is obtained back in 1992 [22] for 15 × 15 multiplexer having 87 GHz output channel spacing using InP technology. A crosstalk of -38 dB is obtained for 380 GHz output channel spacing using silicon nitride (SiN) technology [23]. In practice, the crosstalk can further be improved using high resolution optical lithography [24] along with a repeatable custom fabrication run. Additionally, Although the compact footprint and single chip solution would be lost one could use proven doped $SiO_2$ technology.

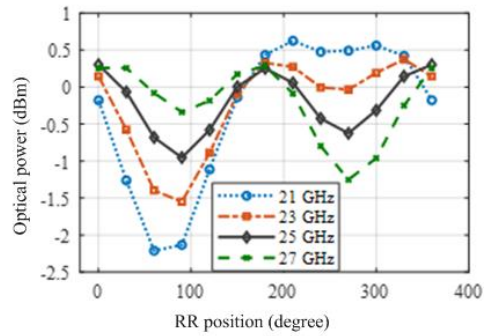

Fig. 13. Simulated optical power as a function of ring resonance position for different frequency shifts between AWG$_1$ and AWG$_2$. The passband width of the AWG output channel used is 20 GHz.

Finally, the simulation results obtained using components designed for fabrication bode well for sensing the whole C-band with high resolution bandwidth albeit with some ripple present in the detectable optical power which may be corrected, if required, by digital signal processing to provide the quantitative spectrum.

## 5. Conclusion

In summary, a simple circuit architecture for on-chip spectral monitor with high resolution is presented. The newly proposed signal processing method and feasibility for photonic integration places the spectrometer in the forefront of the state of the art. Detailed simulation results are presented using constructor design data. As a main component of the proposed design, the integration feasibility of a high-resolution (∼1.30 GHz) ring resonator is demonstrated experimentally. Full tuning over its FSR is achieved. The CMOS compatible $Si_3N_4$ platform is selected for fabrication due to its low loss and maturity. Circuit simulation verify that the spectrum monitoring can be performed over the entire C-band with peak power flatness of ∼1→1.25 dB. The incoherent summation architecture proposed herein is less compact but is more robust to fabrication process variations than the coherent summation architecture proposed previously. A refinement of the architecture is proposed to provide a completely flat spectral response with zero adjacent channel crosstalk at least in theory. Finally, a novel circuit architecture using the combination of an optical switch and multi-input AWG is proposed as a practical more compact alternative implementation.

**Funding**
Huawei Canada sponsored research agreement 'Research on ultra-high resolution on-chip spectrometer'.

**Acknowledgments**

Mehedi Hasan acknowledges the Natural Sciences and Engineering Research Council of Canada (NSERC) for their support through the Vanier Canada Graduate Scholarship program. Trevor J. Hall is grateful to Huawei, Canada for their support of this work. Trevor J. Hall is also grateful to the University of Ottawa for their support of a University Research Chair.

**Disclosures**
The authors declare no conflicts of interest.